# Okun's law revisited. Is there structural unemployment in developed countries?

Ivan O. Kitov
Institute for the Dynamics of the Geopsheres, Russian Academy of Sciences

**Abstract**
Okun's law for the biggest developed countries is re-estimated using the most recent data on real GDP per capita and the rate of unemployment. Our results show that the change in unemployment rate can be predicted with a high accuracy. The link needs the introduction of a structural break which might be caused by the change in monetary policy or/and in measurement units. Statistically, the link between the studied variables is characterized by the coefficient of determination between 0.40 (Australia) and 0.84 (the USA). The residual errors can be associated with measurement errors. The obtained results suggest the absence of structural unemployment in the studied developed countries.

**Introduction**
The Sveriges Riksbank Prize in Economic Sciences in Memory of Alfred Nobel 2010 was awarded to P. Diamond, D. Mortensen and C. Pissarides "for their analysis of markets with search frictions". The core result of their study was an explanation of labour market dynamics including unemployment (e.g. Diamond, 2011; Mortensen and Nagypal, 2007; Pissarides, 2000). Hence, the dynamics of unemployment is a very important and actual problem for the modern economics.

One of the most actively discussed topics related to unemployment is its high persistence since the start of the financial crisis. In the United States, the current rate unemployment is above 9% and it does not show any sign of reduction in the future. There is an opinion that the current situation might manifest tangible structural changes in the labour market. This implies some major changes in the overall organization of the economy when significant parts of it become unnecessary.

We address the problem of structural unemployment by modelling the rate of unemployment using the relationship explaining the dynamics of unemployment by its negative correlation with the growth in GDP – Okun's law (1962). This relation was revisited many times in the past (e.g. Altig, Fitzgerald and P. Rupert, 1997; Knotek, 2007; Tillman, 2010).

We also revisit Okun's law using the most recent data on GDP per capita provided by the Conference Board (2011) and data on unemployment from the OECD (2011). To improve the agreement between the change in unemployment rate and real GDP per capita we introduce structural breaks in Okun's law. Such breaks might manifest artificial changes in definitions of unemployment and real GDP as well as actual shifts in the linear relationship.

We have assessed Okun's law in the biggest developed countries: the United States, France, the United Kingdom, Australia, Canada and Spain. Our results suggest the absence of structural unemployment in the studied developed countries. The persistence of high unemployment is completely related to low rate of real economic growth. In all studied countries, the rate of growth above 2% per year will result in a fall of the unemployment rate

**Okun's law and empirical results**
According to the original form of Okun's law, there exists a negative relation between the growth rate of real GDP and the change in unemployment rate, $du=u_i-u_{i-1}$. The overall GDP includes the change in population as an extensive component which is not necessary dependent on other macroeconomic variables. Econometrically, it is mandatory to use macroeconomic variables of the same origin and we use real GDP per capita, *G*. It is better related to the portion of labor force without job, i.e. the rate of unemployment. Therefore, we rewrite Okun's law in the following form:

$$du = a + bdlnG \qquad (1)$$

where *dlnG=dG/G* is the relative change rate of real GDP per capita, *a* and *b* are empirical coefficients. Okun's law suggests that *b*<0.

We start with the United States and have to introduce a structural break in 1984 into the link. The following relationship was obtained:

$$du = -0.42dlnG + 1.07, t<1985$$
$$du = -0.62dlnG + 1.09, t>1984 \qquad (2)$$

where *dlnG* in the annual growth rate of real GDP per capita, *du* is the annual increment in the rate of unemployment, *u*. Figure 1 displays the observed and predicted *du* between 1958 and 2010. The agreement is excellent. Figure 2 presents some regression results for the curves in Figure 1 with the coefficient of determination $R^2$=0.84. Therefore, more than 84% of the variability in the change of unemployment rate in the U.S. is explained by the change in real GDP per capita. Considering the fact that both macroeconomic variables are measured with an accuracy of approximately 1 percentage point the residual 16% of the variability can be easily associated with the uncertainty in their measurements. Figure 3 demonstrates that the residual error is rather random and does not contain a unit root and has no significant autocorrelation.

Relationship (2) shows that the sensitivity of the *du* to *dlnG* becomes higher after 1984 with the slope of -0.62 and the intercept +1.09. There are two assumptions on the reasons behind this structural break. One is related to the changes in monetary policy in the early 1980s to overcome extremely high inflation. On the other hand, the measures of the GDP deflator and CPI start to deviate around 1980 and the rate of unemployment obtained a new definition in 1984. Thus, the shift in 1984 might also be associated with new units of measurements. In any case, the period after 1984 is described with a very high accuracy including three episodes of unemployment surge in 1991, 2001 and 2009. Moreover, relationship (2) provides a smooth transition through 1984 and describes the fall in unemployment in 1984.

We have also checked several macroeconomic variables as a predictor variable in Okun's law: the overall GDP, GDP per capita corrected for the difference between the whole population and working age population, and productivity as expressed by real GDP per worker. All these variables are inferior to real GDP per capita and thus we use this variable for other countries.

The next country to model is France. We have reversed (1) and obtained the following relationship for *dlnG*:

$$dlnG = -5.0du + 4.6, \quad t<1987$$
$$dlnG = -1.5du + 1.4, \quad t>1986 \qquad (3)$$

Figure 4 present the observed and predicted curves. There is a shift in the dependence around 1987 and the slope in (3) fell from -5.0 to -1.5. Correspondingly, the sensitivity of the unemployment rate to real GDP growth (the reciprocal value of the slope in (3)) increased from -0.2 to -0.67. In order to decrease the rate of unemployment in France, real GDP per capita has to grow by 1.5% per year. When the growth rate is lower, the rate of unemployment increases.

For France, both variables are characterized by an elevated volatility and thus the coefficient of determination $R^2$=0.53 is relatively low. To remove the measurement noise we have smoothed both curves with MA(3). The lower panel in Figure 4 shows that the agreement between the measured increase in real GDP per capita and that predicted from the change in unemployment rate became extremely good.

The United Kingdom also shows an excellent result for Okun's law. The following equation:
$$dlnG = -1.5du + 2.5, \quad t<1987$$
$$dlnG = -2.0du + 1.7, \quad t>1986 \qquad (4)$$

describes the period after 1972. Figure 5 illustrates the agreement between the observed and predicted time series. Before 1972, the OECD provides no unemployment estimates. The year of structural break is the same as in France but the change in the slope is much lower. In any case, to reduce the current rate of unemployment the UK needs to grow at a rate above 1.7% per year in term of real GDP per capita.

For Canada, the following equation was estimated with a structural break around 1985:

$$dlnG = -2.7du + 3.1, \; t<1985$$
$$dlnG = -2.7du + 1.2, \; t>1984 \tag{5}$$

Figure 6 depicts the measured and predicted curves for *dlnG*. For Canada, the rate of growth above 1.2% per year is enough to reduce the rate of unemployment from its current level of 8%. In 2009, $dlnG=-0.033 \; y^{-1}$ and the rate of unemployment rose by 4.7%. In 2010, the change rate of real GDP per capita was $+0.021y^{-1}$ and the rate of unemployment fell by 0.3%.

For Australia, the following equation was estimated with a structural break around 1995:

$$dlnG = -1.7du + 2.4, \; t<1995$$
$$dlnG = -3.0du + 1.2, \; t>1994 \tag{5}$$

The timing of this break is different from those obtained before but the shift in the slope and intercept is big enough to consider it with confidence. It is likely that the structural break was induced by the introduction of a new monetary policy in 1994 (RBA, 1994).

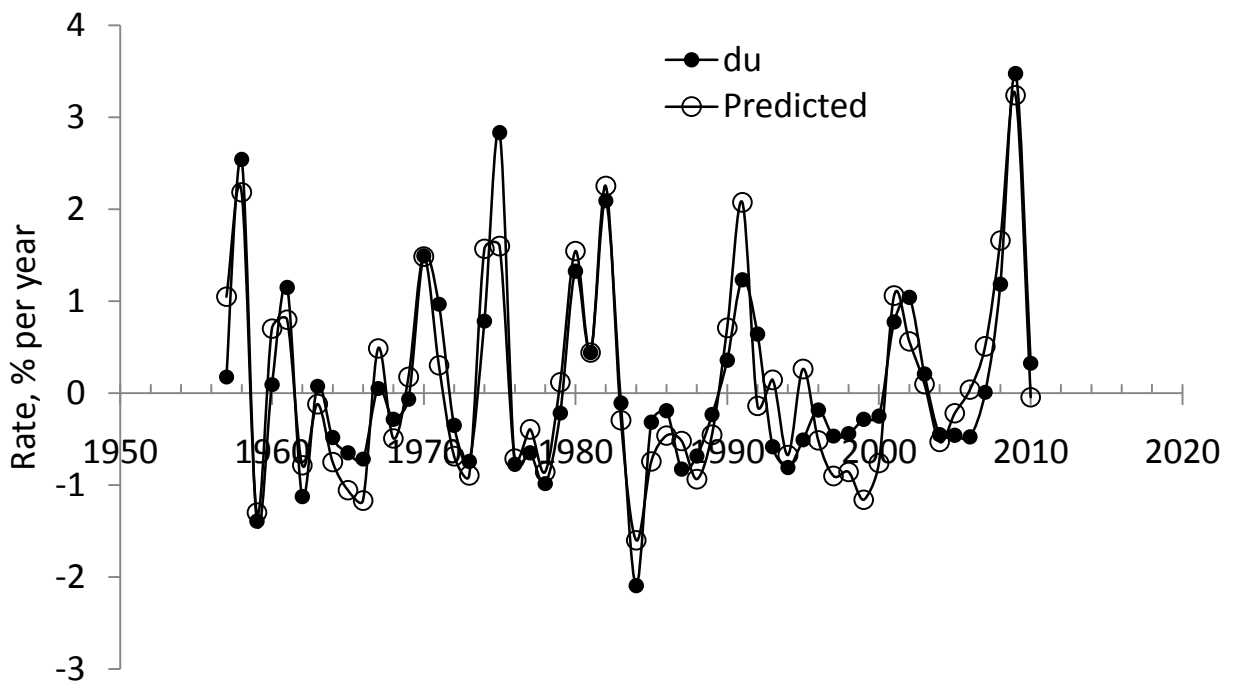

Figure 1. The link between the *du* and *dlnG* in the U.S. as described by relationship (2) with a structural break in 1984

Figure 7 displays the measured and predicted curves for *dlnG*. The overall agreement is not good with $R^2=0.40$ for the period between 1968 and 2010. This low correlation coefficient is associated with the high volatility in both time series. When smoothed with MA(3), the curves in Figure 7 show a much better resemblance. For Australia, the rate of growth above 1.2% per year is enough to reduce the rate of unemployment from its current level of 5.2% (2010). It is similar to Canada.

Finally, the case of Spain is a decisive one. The rate of unemployment in Spain is extremely high and has been varying in a wider range since the 1960s. This is a big challenge for Okun's law. We have obtained the following relationship with a structural break in 1987:

$$dlnG = -2.0du + 5.0, \; t<1987$$
$$dlnG = -0.8du + 2.1, \; t>1986 \tag{5}$$

The timing of this break practically coincides with those in other countries. The sensitivity of unemployment to real economic growth rose significantly in 1987.

Figure 8 displays the measured and predicted curves for *dlnG*. The overall agreement is not good with $R^2=0.59$ for the period between 1961 and 2010. When smoothed with MA(3), the curves in Figure 8 show an extraordinary agreement but one can also introduce another structural break near 1968.

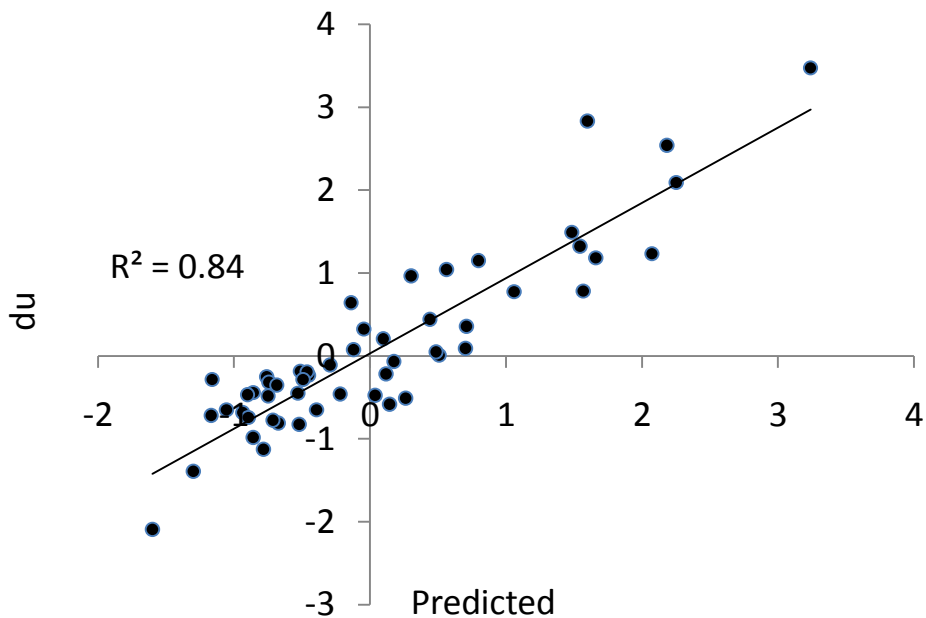

Figure 2. A scatter plot *du* against *dlnG* from Figure 1 with a linear regression line. $R^2=0.84$.

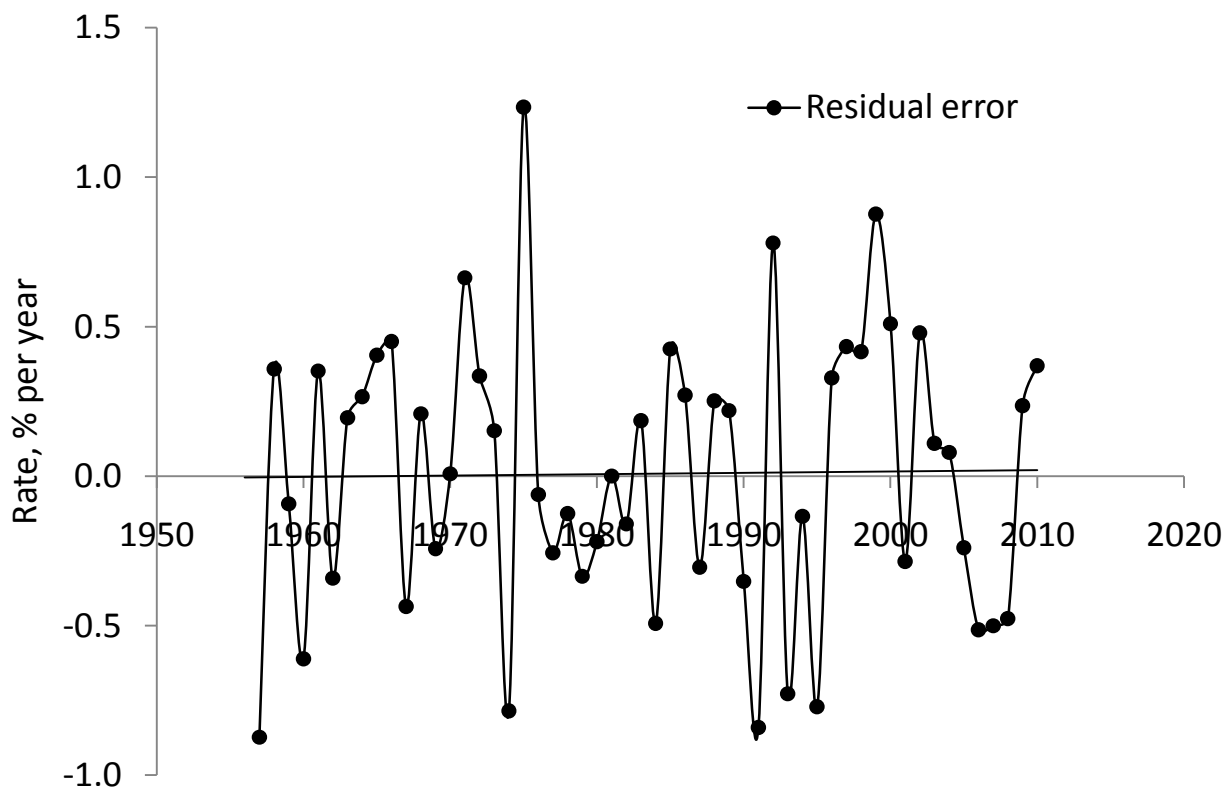

Figure 3. The model error for the U.S.

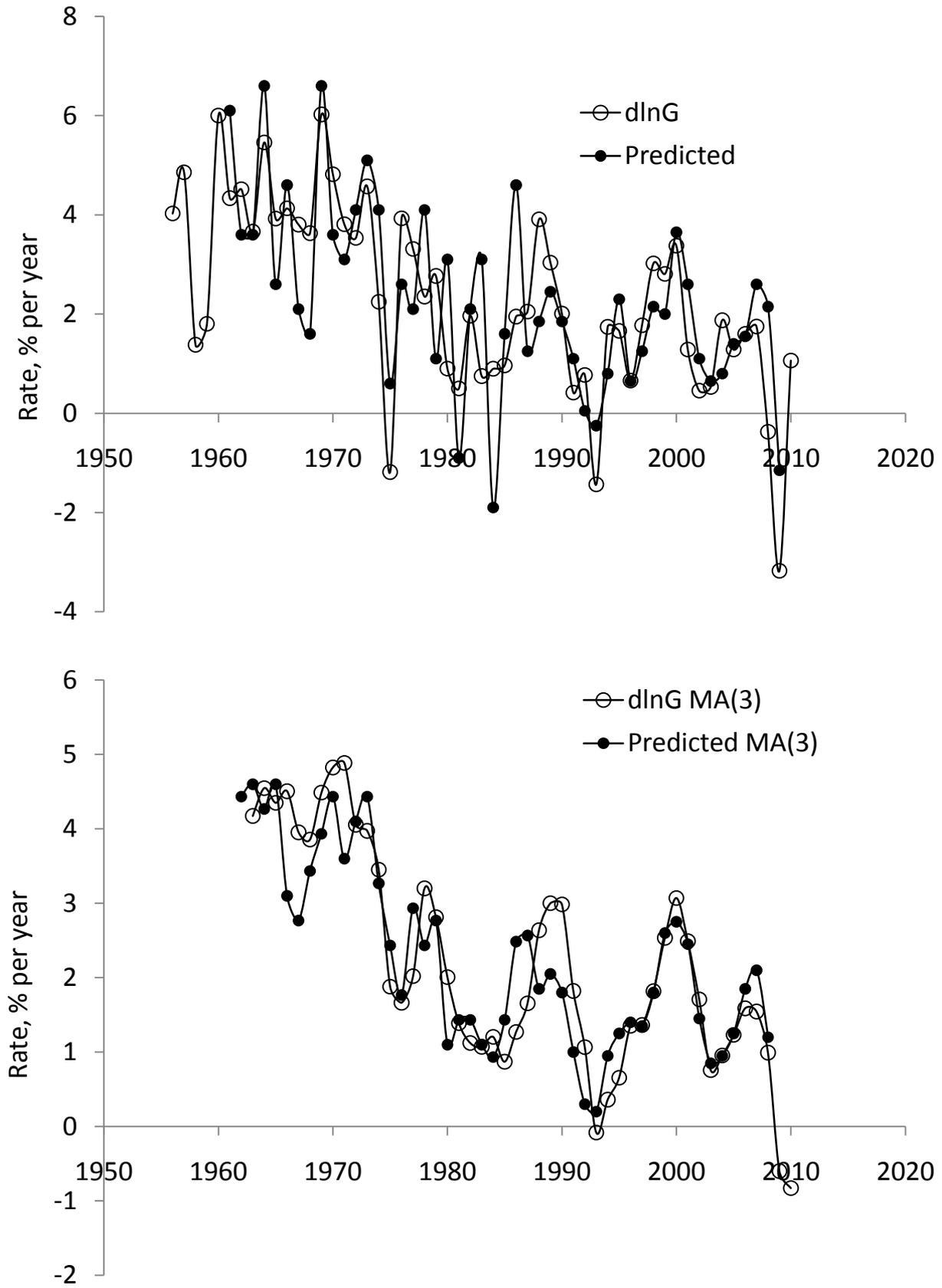

Figure 4. Observed and predicted *dlnG* for France. The lower panel presents the curves smoothed with MA(3).

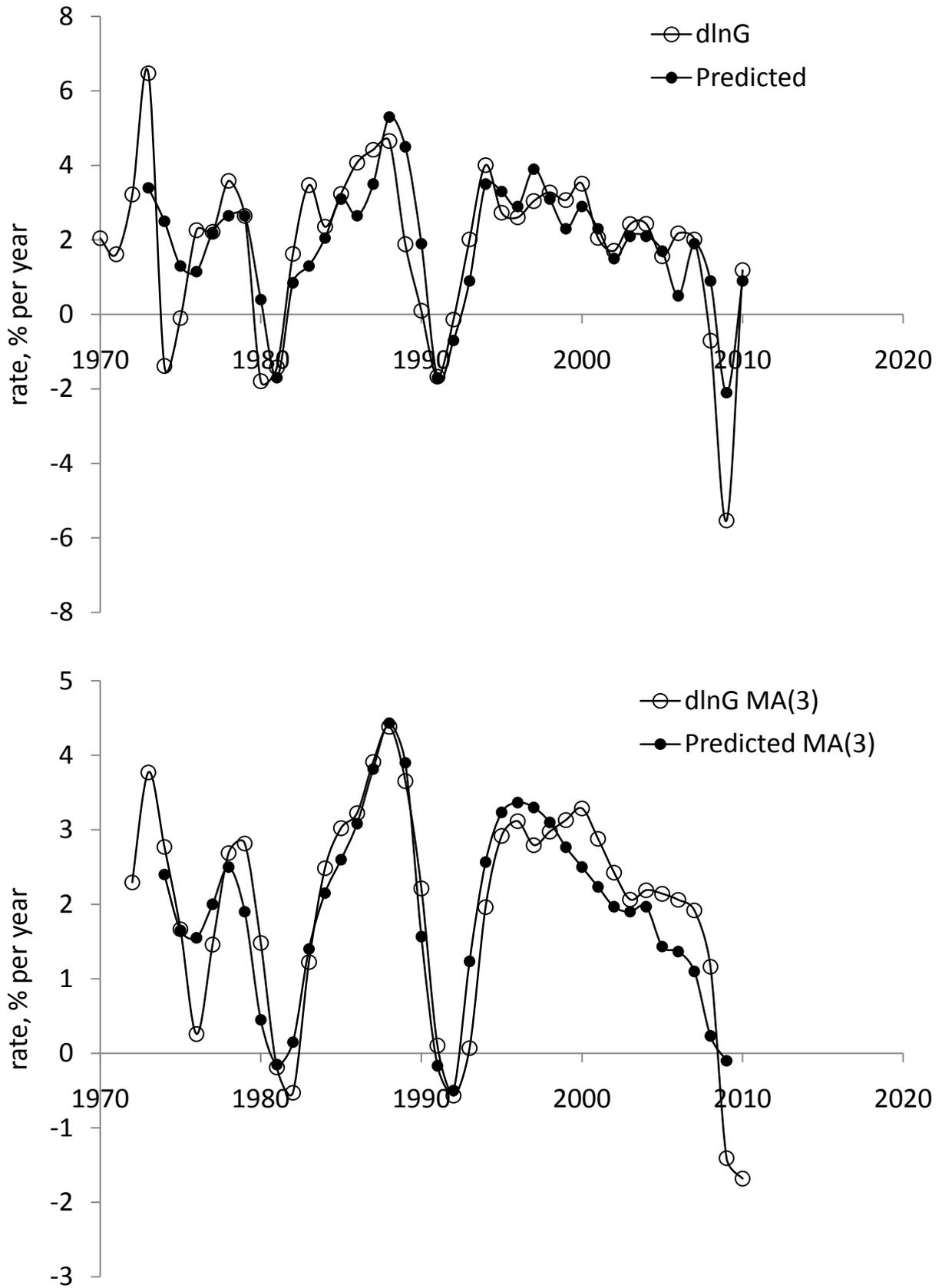

Figure 5. Observed and predicted *dlnG* for the UK. The lower panel presents the curves smoothed with MA(3).

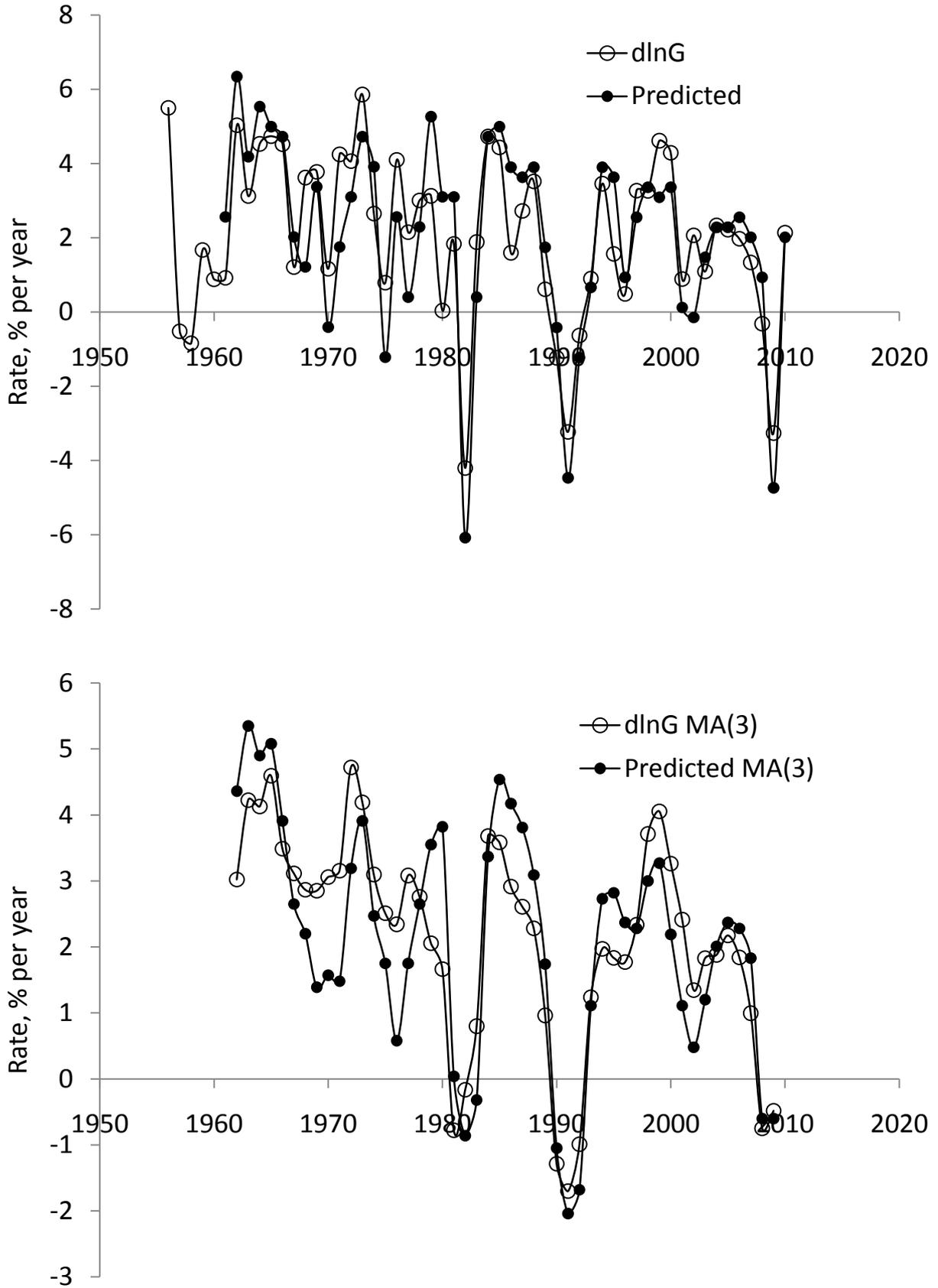

Figure 6. Observed and predicted *dlnG* in Canada. The lower panel presents the curves smoothed with MA(3).

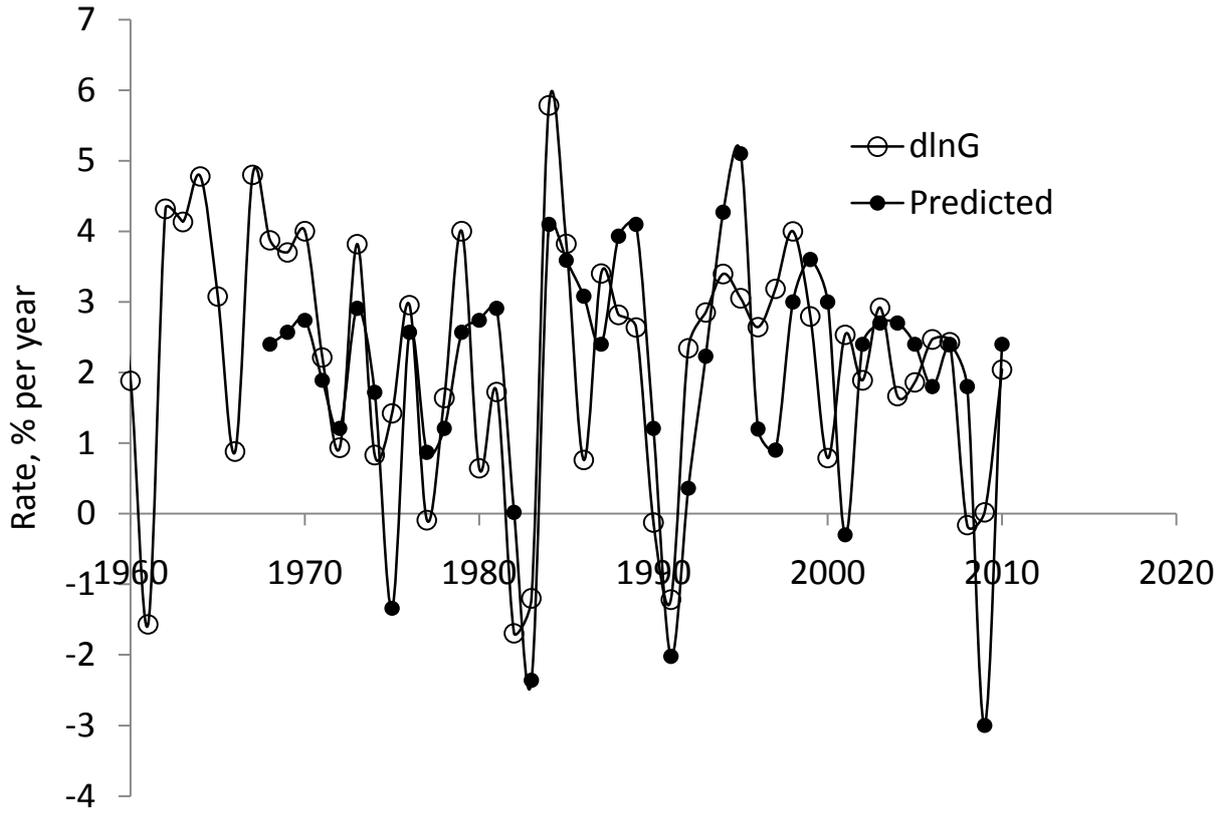
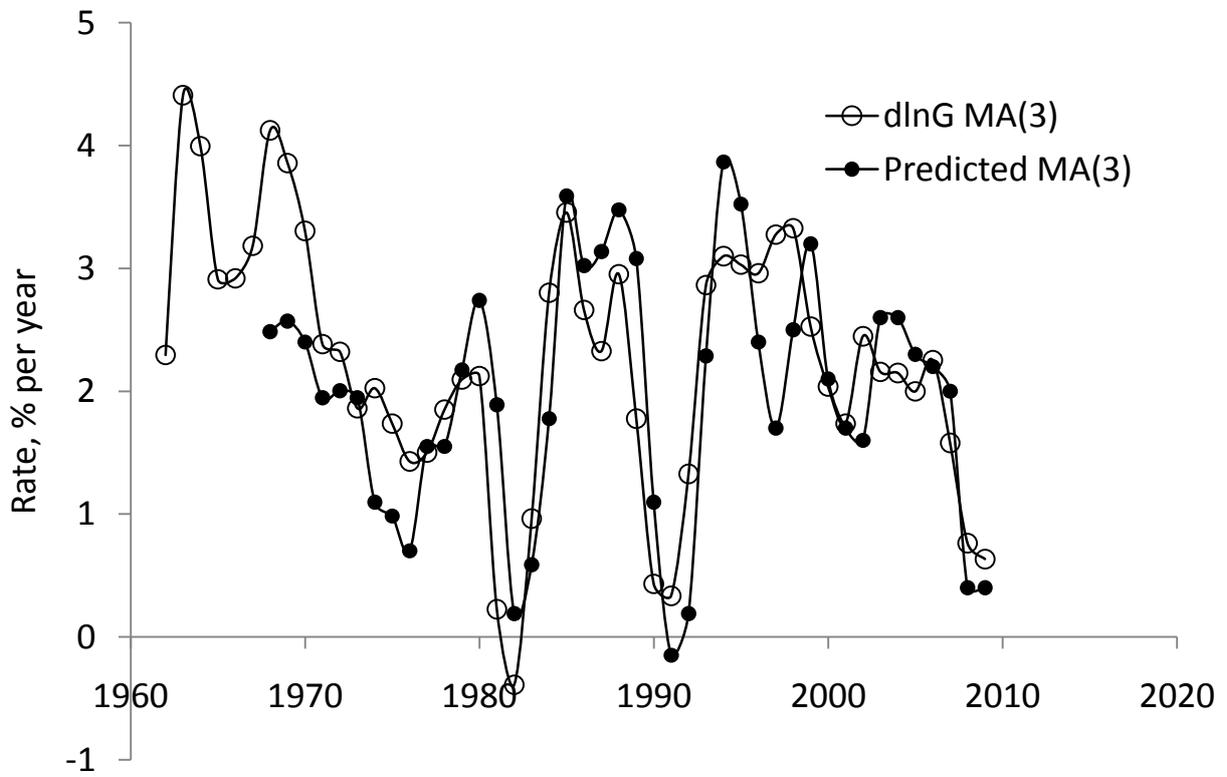

Figure 7. Observed and predicted *dlnG* in Australia. The lower panel presents the curves smoothed with MA(3).

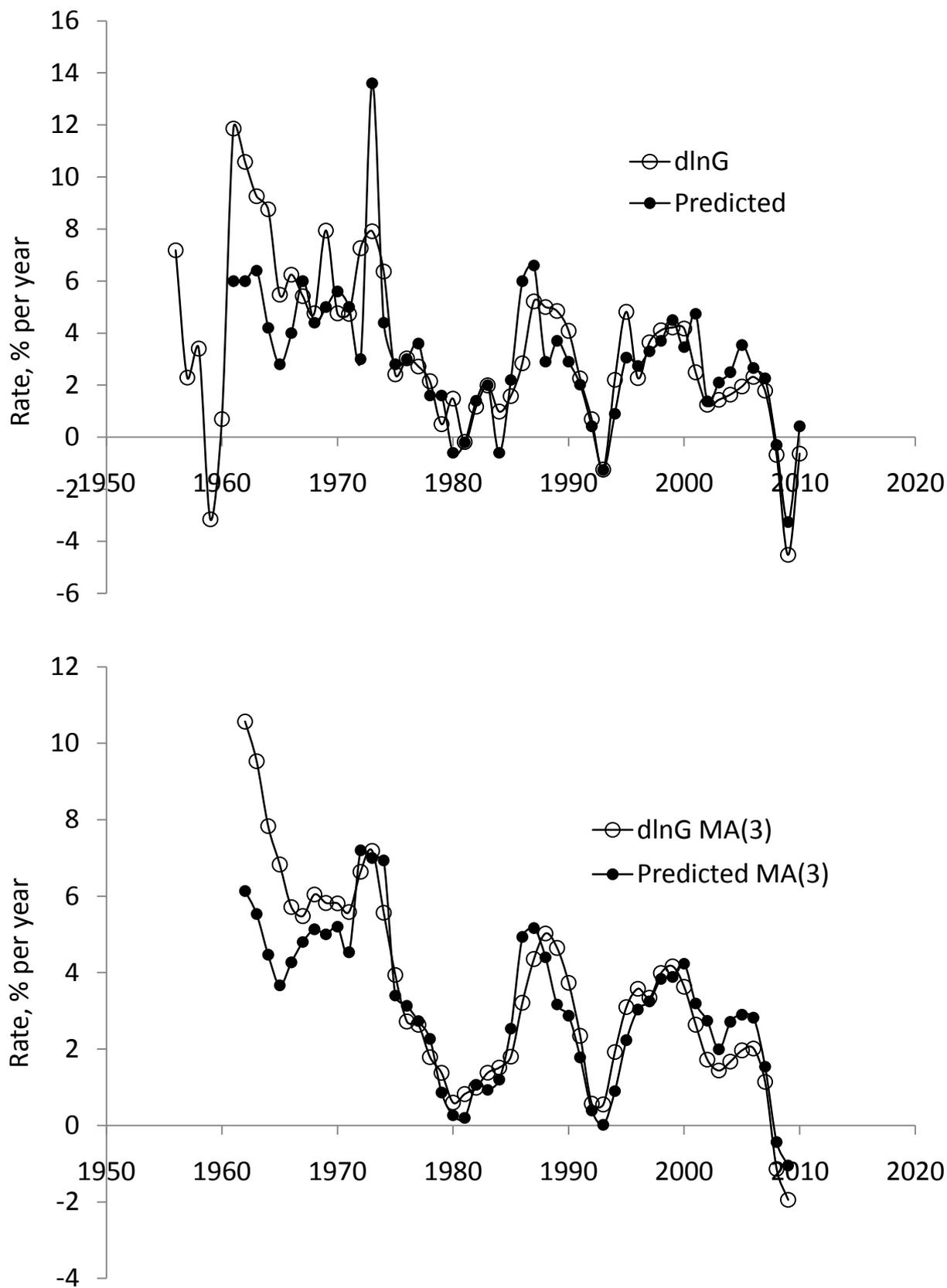

Figure 8. Observed and predicted *dlnG* in Spain. The lower panel presents the curves smoothed with MA(3).

**Conclusion**

With real GDP per capita instead of the overall GDP, Okun's law demonstrates an extraordinary predictive power for the biggest developed countries. One can accurately describe the dynamics of unemployment since the 1960s. The currently high levels of unemployment in developed countries cannot be reduced without fast economic growth well above 2% per year. In that sense, there are no structural unemployment components in the currently high rates of unemployment in the studied countries.